\documentstyle[prl,aps]{revtex}

\begin{document}

\twocolumn[{
\draft
\widetext
\title
{Early times in tunneling }
\author{Gast\'on Garc\'{\i}a-Calder\'on}
\address{{ \it Instituto de F\'{\i}sica,
Universidad Nacional Aut\'onoma de M\'exico\\
Apartado Postal 20-364, 01000 M\'exico, D.F., M\'exico}}
\author{Jorge Villavicencio}
\address{\it {
Facultad de Ciencias, Universidad Aut\'onoma de Baja California\\
Apartado Postal 1880, Ensenada, B.C., M\'exico}}

\date{20 June 2000}

\mediumtext

\begin{abstract}
Exact analytical solutions of the time-dependent Schr\"odinger
equation with the initial condition of an incident cutoff wave
are used to investigate the traversal time for tunneling.
The probability density  starts from a vanishing value
along the tunneling  and transmitted regions of the potential.
At the barrier width it exhibits, at early times, a distribution
of traversal times that typically has a peak $\tau_p$ 
and a width $\Delta \tau$. 
Numerical results for other tunneling times, as the phase-delay time,  
fall within $\Delta \tau$. The B\"uttiker traversal time
is the closest to $\tau_p$.
Our results  resemble calculations based on Feynman paths
if its noisy behaviour is ignored.
\end{abstract}
\maketitle

}]

\narrowtext
Quantum tunneling, that refers to the possibility that a particle
traverses through a classically forbidden region, constitutes one
of the paradigms of quantum mechanics. In the energy domain, where
one solves the stationary Schr\"odinger equation at a fixed energy
$E$, tunneling is well understood.
In the time domain, however, there are aspects still open to
investigation. Recent technological achievements as the possibility
of constructing artificial quantum structures at nanometric
scales\cite{qs} or the manipulation of individual atoms\cite{corral}
have stimulated work on time-dependent tunneling both at applied and
fundamental levels. A problem that has remained controversial
and the subject of a great deal of attention over the years is the
tunneling time problem\cite {traversal}, that can be stated as
the question: How long it takes to a particle to traverse a
classically forbidden region? In time-dependent tunneling, many
works attempting to answer the above question consider the numerical
analysis of the time-dependent Schr\"odinger equation with the initial
condition of a Gaussian wave packet\cite{hartman,collins,muga}.
A common feature of the majority of these approaches
is that the initial wave packet extends through all space. As a
consequence the initial state, although it is manipulated to reduce
as much as possible its value along the tunneling and transmitted
regions, contaminates from the beginning the tunneling process and
hence usually it is required a long time analysis of the solutions.
The above situation may be circumvented by considering cutoff wave 
initial conditions\cite{mm,stevens83,morettipra92,muga96,gcr97}.

In this work we consider a analytic time-dependent solutions
to the Schr\"{o}dinger
equation with the initial  condition at $\tau=0$ of a incident cutoff 
wave, to investigate the traversal time for tunneling through a 
potential barrier. The problem may be visualized as a
{\it gedanken experiment} consisting of a shutter, situated
at $x=0$, that separates a beam of particles from a potential barrier
of height $V_0$ located in the region $0\leq x\leq L$. At $\tau=0$ the
shutter is opened. The probability density rises
initially from a vanishing value and evolves with time through 
$x > 0$. At the barrier edge $x=L$, the probability density at
time $\tau$, yields the probability of finding the particle after
a time $\tau$ has elapsed.
Since initially there is no particle along the tunneling region,
detecting the particle at the barrier edge at time $\tau$, 
provides a measure of its traversal time through the tunneling region.

The transient behavior of the time-dependent
solution at early times and at distances close to the interaction 
region plays a significant role in our approach.
Other formulations, based on the stationary solutions of the
Schr\"{o}dinger equation\cite{hartman,smith}, refer 
to asymptotically long times at large distances and hence ignore
transient effects. These approaches provide a
single value for the traversal time. In contrast,
our approach leads to a distribution of traversal times 
as in works based on the Feynman path integral
method\cite{sokolovski,fertig,yamada}, though as indicated below
both approaches differ in important aspects.

In a recent paper we have obtained the time-dependent solution to the
Schr\"{o}dinger equation for tunneling through an arbitrary potential
of finite range with the initial condition of a cutoff plane wave of
momentum $k$. The solution may be written as a term proportional to the
free solution plus a contribution involving an infinite sum of resonance
terms associated with the $S-$matrix poles of the potential\cite{gcr97}.
Our approach is based on the Laplace transform technique and considers
some analytical properties of the outgoing wave propagator. Some
decades ago Moshinsky considered the free case solution to the above
problem\cite{mm}. Moshinsky showed that the probability density,
for a fixed value of the distance $x_0$ as a function
of $t$, exhibits a transient regime that he named diffraction in time.
Recently, observations of that phenomenon have been reported\cite{exp}.
For the sake of simplicity in our approach, as Moshinsky also did,
we consider the instantaneous removal of the shutter.
This may be seen as a kind of `sudden approximation' to a shutter
opening with finite velocity, where the treatment becomes more
involved\cite{gahler}. As shown below the terms depending on the
$S-$matrix poles provide a novel transient behavior that may dominate
the early times in the tunneling process.

The plane wave cutoff initial condition discussed in
Refs. \cite{mm,gcr97} refers to a shutter that acts as a perfect
absorber (no reflected wave). One can also envisage a shutter that
acts as a perfect reflector. In such a case the initial
wave may be written as, \begin{equation}
\psi _s(x,k,\tau =0)=\left\{ 
\begin{array}{cc}
e^{ikx}-e^{-ikx}, & x<0 \\ 
0, & x>0.
\end{array}
\right.   
\label{2a}
\end{equation}
One can then proceed along lines similar to those discussed in
Ref.\ \cite {gcr97} to derive the time-dependent solution
$\psi _s(x,k,\tau )$ of the Schr\"{o}dinger equation for a potential
$V(x)$ that vanishes outside the region $0\leq x\leq L$. The
solution along the internal region reads,
\begin{eqnarray}
\psi _s(x,k,\tau ) &=&\phi (x,k)M(0,k,\tau )-\phi (x,-k)M(0,-k,\tau)
\nonumber \\
&&-\sum_n^\infty \phi _n(x)M(0,k_n,\tau ),\,\,\,\,(0\leq x\leq L)
\label{3b}
\end{eqnarray}
where $\phi (x,k)$ refers to the stationary solution and
$\phi_n(x)=2iku_n(0)u_n(x)/(k^2-k_n^2)$ is given in terms of
the resonant (Gamow) states $\{u_n(x)\}$ and complex poles $\{k_n\}$
of the problem\cite{gcr97,gcp76}.
Similarly the transmitted solution\cite{foot} becomes,
\begin{eqnarray}
\psi _s(x,k,\tau ) &=&T(k)M(x,k,\tau )-T(-k)M(x,-k,\tau )
\nonumber \\
&&-\sum_n^\infty T_nM(x,k_n,\tau ),\,\,\,\,(x\geq L)
\label{3c}
\end{eqnarray}
where $T(k)$ and $T(-k)$ are transmission amplitudes, and
$T_n=2iku_n(0)u_n(L){\rm exp}(-ik_nL)/(k^2-k_n^2)$. In the
above two equations the functions $M(x,k,\tau )$ and
$M(x,k_n,\tau)$ are defined as,
\begin{equation}
M(x,q,t)=\frac 12{\rm e}^{(imx^2/2\hbar \tau )}
{\rm e}^{y_q^2}{\rm erfc} (y_q),
\label{4}
\end{equation}
where the argument $y_q$ is given by 
\begin{equation}
y_q\equiv {\rm e}^{-i\pi /4}\left( \frac m{2\hbar \tau}
\right) ^{1/2}\left[x-\frac{\hbar q}m\tau \right].
\label{5}
\end{equation}
In Eqs. (\ref{4}) and  (\ref{5}) $q$ stands either for $k$ or
$k_n$, the index $n$ refers to a given complex pole.
Poles are located on the third and
fourth quadrants of the complex $k$-plane. The solution for the
free case with the reflecting initial condition is,
$\psi _s^0(x,k,\tau )=M(x,k,\tau )-M(x,-k,\tau )$.
>From the analysis given in Ref.\ \cite{gcr97} one can see that the
above exact solutions satisfy the corresponding initial conditions,
i.e., they vanish exactly for $x>0$. At very long times it is also
shown in Ref.\ \cite{gcr97} that the terms $M(x,k_n,\tau )$ that
appear in the above equations vanish. The same occurs for
$M(x,-k,\tau )$ while, as shown firstly in Ref. \cite{mm},
$M(x,k,\tau )$ tends to the stationary solution. Hence, at long
times, each of the above exact solutions go into the corresponding
stationary solutions, namely, along the internal region as
$\psi (x,\tau)=\phi (x,k){\rm exp}(-iE\tau /\hbar )$ and along
the external region as $\psi (x,\tau )=T(k){\rm exp}(ikx)
{\rm exp}(-iE\tau /\hbar )$. Note that  at early times and short
distances there is a competition between the contribution of the
free-type terms ($M$ functions depending on $k$) and the pole terms
($M$ functions depending on either $k_n$ or $k_{-n}$)
in Eqs. (\ref{3b}) and (\ref{3c}).
As exemplified below, depending
on the potential parameters one may have the predominance of one or
the other type of terms. Note also that the initial state is not
strictly monochromatic (it extends from $-\infty $ to $0$) and hence
it has a distribution of components around $k$ in momentum space.
One could construct an initial cutoff wavepacket as a linear
combination of cutoff waves. However,
since we compare below with definitions of tunneling times
involving plane waves, wavepackets will not be considered here.
Besides, in general they involve no negligible momentum components
above the barrier potential and hence obscure the dynamics of
tunneling.

In order to apply the above ideas, we consider a model that has been
used extensively for the tunneling time problem, namely,
the rectangular barrier potential, characterized by a height $V_0$
in the region $0\leq x\leq L$. The shutter is located at $x=0$.
In order to calculate Eqs.\ (\ref{3b}) and (\ref{3c}) for the initial
condition (\ref{2a}), in addition to the parameters $V_0$, $L$, and that
corresponding to the incident energy $E=\hbar ^2k^2/2m$, we need to
determine the complex poles $\{k_n\}$ and resonant states
$\{u_n(x)\}$. It is well known that for a finite range potential
there are an infinite number of poles. The $S-$matrix poles for
the rectangular barrier potential may be obtained from the corresponding
transmission amplitude $T(k)=4kq{\rm exp}(-ikL)/J(k)$, where
$q=[k^2-k_0^2]^{1/2}$ with $k_0^2=2mV_0/\hbar ^2$. They correspond to
the  zeros of $J(k)$ in the $k-$plane, namely,
\begin{equation}
J(k)=(q+k)^2{\rm exp}(-iqL)-(q-k)^2{\rm exp}(iqL)=0.
\label{6}
\end{equation}
We follow a well established method to obtain the solutions to the above
equation\cite{gcr97,nussenzveig}. The resonant states of the problem
satisfy the time-independent Schr\"{o}dinger equation of the problem
with outgoing boundary conditions\cite{gcr97}. They read,
\begin{equation}
u_n(x)=C_n\left[ {\rm e}^{iq_nx}+b_n{\rm e}^{-iq_nx}\right]
,\,\,\,(0\leq x\leq L)
\label{7}
\end{equation}
where $b_n=(q_n+k_n)/(q_n-k_n)$ and $C_n$ may be obtained from the
normalization condition\cite{gcr97}, 
\begin{equation}
\int_0^Lu_n^2(x)dx+i{\frac{u_n^2(0)+u_n^2(L)}{2k_n}}=1.
\label{7a}
\end{equation}
Note that both the complex poles and the resonant states are a
function of $V_0$ and $L$ and hence are a property of the system.

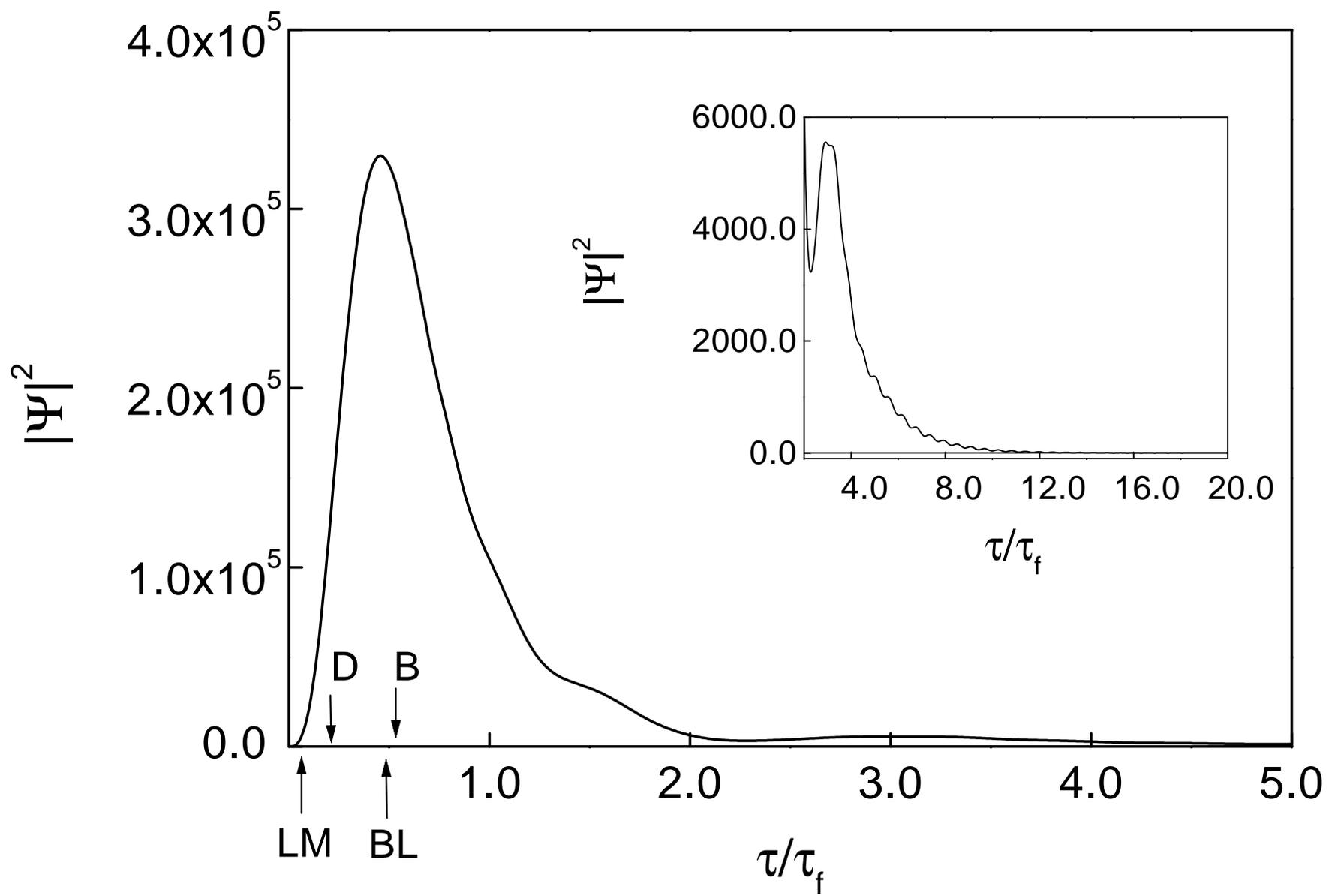
\begin{figure}[tbp]
\caption{Plot of $|\psi(L,\tau)|^2$ at the barrier edge $x=L=10\, nm$
as a function of time in units of the free passage time $\tau_f$.
The inset shows $|\psi(L,\tau)|^2$ at larger times. The arrows indicate
the values of the Larmor time LM, the semi-classical time BL, the
B\"uttiker traversal time B, and the phase-delay time D. See text.}
\label{fig1}
\end{figure}

To exemplify the time evolution of the probability density we consider
the set of parameters: $V_0=0.711\,eV$, $L=10\,nm$, $E=0.1422\,eV$,
$m^{*}=0.067\,m_e$, inspired in semiconductor quantum
structures\cite{qs}.
Our choice of parameters guarantees that most momentum components of
the initial state tunnel through the potential. Figure \ref{fig1}
shows a plot of $|\psi(L,\tau )|^2$, calculated at the barrier
edge $x=L$ as a function of time in units of
the free passage time $\tau_f=mL/(\hbar k)=11.56\,fs$.
We have used Eq.\ (\ref{3c}), though Eq.\ (\ref{3b}) holds the same.
The time-dependent solution is normalized by
$|T(k)|^2=5.332\times 10^{-9}$.
One sees that as soon as
$\tau \neq 0$ the probability density starts to grow up. As discussed
elsewhere\cite{gcrv99}, this is due to the non-relativistic
character of the description.
Einstein causality may be fulfilled by cutting off the contributions
to the probability density smaller than $\tau _0=L/c$. In our example
$\tau_0=0.033\,fs$ or $\tau_0/\tau_f =0.0028$, too small to
be appreciated in Fig. \ref{fig1}.
At early times one sees a time domain resonance structure. Thereafter
the probability density approaches essentially its asymptotic value. 
We found that the
resonant sum is the relevant contribution to the time domain resonance
since that of the free-type term is quite small and varies smoothly
with time. In the transmitted region, $x>L$, not shown here, the
time domain resonance becomes a propagating structure, as
follows from Eq.\ (\ref{3c}). The time domain resonance
corresponds to a transient effect and as it propagates through the
transmitted region becomes smaller and smaller. Asymptotically, at
large distances and times, it becomes very small while the free-type
term becomes the dominant contribution with its wavefront propagating 
with velocity $v=\hbar k/m$.
Calculations using the absorbing initial condition exhibit a similar
time domain resonance. Hence a linear combination of reflecting and
absorbing initial conditions should also exhibit it.

The time domain resonance represents a distribution of traversal times.
The corresponding peak represents the largest probability to find the
tunneling particle at the barrier edge $x=L$. In our example, as
shown in Fig. \ref{fig1}, the time domain resonance peaks at
$\tau _p=5.326\,fs$, faster than the free passage time
across the same distance of $10$ $nm$, that is,
$\tau _p/\tau _f=0.46$.
Note that the distribution is quite asymmetric. Altough the first
resonance term of the solution provides the main contribution, 
convergence of the series usually requires to sum up to $100$ terms.
The inset displays the probability density from $\tau /\tau _f=2$ 
up to $\tau /\tau _f=20$. One sees a small
structure around $\tau /\tau _f=3$ and then the probability
density decreases very fast towards unity, the stationary
regime.
The main range of traversal times 
occurs around the peak value $\tau_p$. We define the width 
of the distribution, $\Delta \tau$, by the rule of the half-width 
at half-maximium. This yields $\Delta \tau = 13.48\,fs$ or
$\Delta \tau/\tau_f = 1.16$. The resonance is broad, since 
$\Delta \tau \approx 2 \tau_p$.
We have found that for fixed $V_0$ 
and $E$, and a decreasing $L$, the width diminishes. The same occurs for
fixed $E$ and $L$, and an increasing $V_0$. Systematically, however,
$\Delta \tau > \tau_p$.
For the sake of comparison, the arrows
in Fig. \ref{fig1} indicate the values calculated for a number of
definitions of tunneling times existing in the literature for the
rectangular barrier potential\cite{buttiker}, as the Larmor time of
Ba\'{z} and Rybachenko, $\tau _{LM}$; the semi-classical or
B\"{u}ttiker-Landauer time, $\tau _{BL}$;
the B\"{u}ttiker traversal time, $\tau _B$, and the phase-delay time,
$\tau_D$ \cite{eqs}. All of them fall within the broad range of values
given by $\Delta \tau$. Note, however, that the B\"{u}ttiker traversal 
time $\tau _B$ is the closest to $\tau_p$.
As shown below we have found this situation extensively in our
numerical calculations. Also, since the barrier is opaque,
$\tau_B$ is close to $\tau_{BL}$.

We refer briefly to approaches to the tunneling
time problem  based on the Feynman path integral
method\cite{sokolovski,fertig,yamada}. For plane waves and a
rectangular barrier potential Fertig\cite{fertig} has derived an
expression, $C(\tau)$, that gives the probability amplitude
that a particle remains a time $\tau $ in a region, 
(Eq. (3) of Ref. \cite{fertig}). 
Recently Yamada\cite{yamada} has plotted $G(\tau )=|C(\tau )|^2$
versus $\tau $ (Fig. 2 of Ref.\cite{yamada}). His parameters
are the same as in our Fig. \ref{fig1}, {\it i.e.},
$V_0/E=5$ and $kL=5$. Our calculation for
$|\psi(L,\tau )|^2$ resembles the average shape of
$G(\tau )$, provided its noisy behavior is ignored.
Note, however, that the meaning of both quantities
is different. As indicated by Yamada, $G(\tau )$ refers to a
`residence time'\cite{residence} whereas our approach 
corresponds to a `passage' or traversal time\cite{passage}.

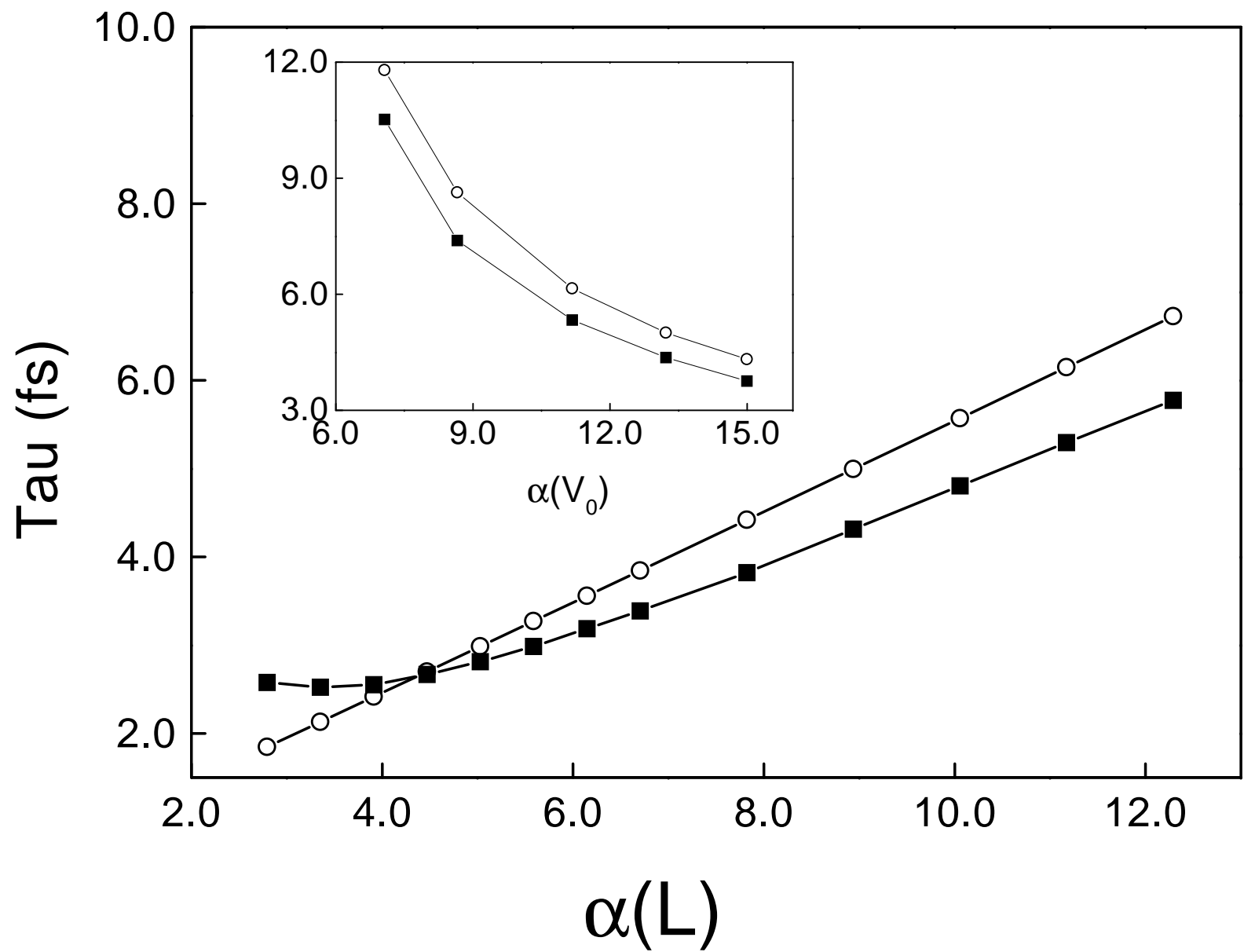
\begin{figure}[tbp]
\caption{Plot of the exact time domain resonance peak $\tau_p$ (solid
squares) versus the opacity $\alpha(L)$ ($V_0$ fixed). For comparison
we plot the B\"uttiker traversal time $\tau_B$ (hollow circles). The
inset shows a similar calculation versus the opacity $\alpha (V_0)$
(L fixed). See text.}
\label{fig2}
\end{figure}

In Fig. \ref{fig2} we plot $\tau _p$ (solid squares)for different
values of the opacity $\alpha =k_0L$, with $k_0=[2mV_0]^{1/2}/\hbar $.
Keeping $V_0$ fixed and varying $L$ defines $\alpha (L)$. We can then
identify two regimes, one in the range $2\leq \alpha (L)\leq 5$, the
tunneling regime, where $\tau _p$ remains almost constant as
$\alpha (L)$ increases, and another regime, the opaque regime, with
$\alpha (L)>5$, where we find that $\tau _p$ increases linearly.
The first behaviour above is related to the first top-barrier S-matrix 
pole and the second one to the components of the incident wave that go 
above the barrier. 
There is still another regime, not shown in Fig. \ref{fig2},
where $\alpha (L)<1$, that corresponds to very shallow or
very thin barriers or both and will not be considered here.
There the free-type terms in Eq. \ (\ref{3c}) dominate over the resonant contribution.
For comparison we plot the B\"{u}ttiker traversal time $\tau _B$
(hollow circles). We see that $\tau _B$ remains rather close to
$\tau _p$. Note, however, that $\tau_B$  behaves linearly in the whole 
range. This different qualitative behaviour as a function of $L$
between both times deserves further study.
The inset in Fig. \ref{fig2}
exhibits a similar comparison for the opacity $\alpha (V_0)$,
with $L$ fixed and varying $V_0$. Here we observe that $\tau _B$
remains quite close to $\tau _p$ in the whole range.
Regarding the phase-delay time $\tau_D$, its predictions 
usually fall within the  width $\Delta \tau$.
For fixed $V_0$ and $E$, $\tau_D$ as a function of 
$L$ exhibits  qualitatively a different behaviour than that 
of Fig. \ref{fig1} (See Fig. 5 in ref. \cite{hartman}).

To end we stress that the largest probability to find the 
tunneling particle at the barrier width, given by $\tau_p$,
is sensitive to both variations of the barrier width $L$ and 
of the height $V_0$, and also, that the B\"uttiker traversal 
time is found very close to the value of $\tau_p$ though we find 
qualitative differences between them as a function of $L$.

G. G-C. thanks M. Moshinsky for useful discussions and acknowledges
support of DGAPA-UNAM under grant IN116398. We also acknowledge
partial financial support of Conacyt under contract no.
431100-5- 32082E.

\begin{references}
\bibitem{qs}  E. E. Mendez, in {\it Physics and Applications of Quantum
Wells and Superlattices}, edited by E. E. Mendez and K. Von Klitzing
(Plenum, New York, 1987) p. 159.

\bibitem{corral}  M. F. Crommie, C. P. Lutz, and D. M. Eigler,
Science {\bf %
262}, 218 (1993).

\bibitem{traversal}  E. H. Hauge and J. A. Stovneng, Rev. Mod. Phys
{\bf 61}, 917 (1989); R. Landauer and Th. Martin,{\it \ ibid.}
{\bf 66}, 217 (1994).

\bibitem{hartman}  T. E. Hartman, J. Appl. Phys. {\bf 33}, 3427 (1962).

\bibitem{smith}  F. Smith, Phys. Rev. {\bf 118}, 349 (1960).

\bibitem{sokolovski}  See for example: D. Sokolovski, S. Brouard,
and J. N. L. Connor, Phys. Rev. A {\bf 50}, 1240 (1994).

\bibitem{fertig}  H. A Fertig, Phys. Rev. Lett. {\bf 65}, 2321 (1990);
Phys. Rev. B {\bf 47}, 1346 (1993).

\bibitem{yamada}  N. Yamada, Phys. Rev. Lett. {\bf 83}, 3350 (1999).

\bibitem{collins}  S. Collins, D. Lowe, and J. R. Barker, J. Phys. C
{\bf 20}, 6213 (1987).

\bibitem{muga}  V. Delgado and J. G. Muga, Ann. Phys. (N.Y) {\bf 248},
122 (1996).

\bibitem{mm}  M. Moshinsky, Phys. Rev. {\bf 88}, 625 (1952).

\bibitem{stevens83}  K. W. H. Stevens. J. Phys. C {\bf 16}, 3649 (1983).

\bibitem{morettipra92}  P. Moretti, Phys. Rev. A {\bf 46}, 1233 (1992).

\bibitem{muga96}  S. Brouard and J. G. Muga, Phys. Rev. A {\bf 54},
3055 (1996).

\bibitem{foot}  For the rectangular barrier, an
expression analogous to Eq. (3) may be derived without using 
resonant states, G. Garc\'{\i }a-Calder\'{o}n, J. L. Mateos
and M. Moshinsky (unpublished).

\bibitem{gcr97}  G. Garc\'{\i }a-Calder\'{o}n and A. Rubio, Phys.
Rev. A {\bf 55}, 3361 (1997).

\bibitem{exp}  P. Szriftgiser, D. Gu\'{e}ry-Odelin, M. Arndt, and J.
Dalibard, Phys. Rev. Lett. {\bf 77}, 4 (1996); Th. Hils, J. Felber, R.
G\"{a}hler, W. Gl\"{a}ser, R. Golub, K. Habicht, and P. Wille, Phys.
Rev. A {\bf 58}, 4784 (1998).

\bibitem{gahler}  R. G\"{a}hler and R. Golub, Z. Phys. B {\bf 56},
5 (1984).

\bibitem{gcp76}  G. Garc\'{\i }a-Calder\'{o}n and R. E. Peierls,
Nucl. Phys. A {\bf 265}, 443 (1976).

\bibitem{gcrv99}  G. Garc\'{\i }a-Calder\'{o}n, A. Rubio and J.
Villavicencio, Phys. Rev. A {\bf 59}, 1758 (1999).

\bibitem{nussenzveig}  H. M. Nussenzveig, Nucl. Phys. {\bf 11},
499 (1957).

\bibitem{buttiker}  M. B\"{u}ttiker, Phys. Rev. B {\bf 27}, 6178 (1983).

\bibitem{eqs}  See, respectively, Eqs. (1.4), (1.7), (3.12) and (3.2) of
Ref. \cite{buttiker}.

\bibitem{residence}  The `residence time' involves the integral
of the probability density along the internal region of the 
interaction. This definition does not distinguish whether 
particles are finally transmitted or reflected and hence
it is not appropriate to describe traversal times.

\bibitem{passage}  This difference is relevant because 
Yamada has argued, using a 'weak decoherence condition',
that for `residence time' type a probability distribution of 
tunneling times is not definable (see ref. \cite{yamada}). 

\end{references}
\end{document}